\documentclass{aastex63}

\newcommand{\PSUAA}{Department of Astronomy \& Astrophysics, 525 Davey Laboratory, The Pennsylvania State University, University Park, PA, 16802, USA}
\newcommand{\PSUCEHW}{Center for Exoplanets and Habitable Worlds, 525 Davey Laboratory, The Pennsylvania State University, University Park, PA, 16802, USA}
\newcommand{\PSETI}{Penn State Extraterrestrial Intelligence Center, 525 Davey Laboratory, The Pennsylvania State University, University Park, PA, 16802, USA}

\shorttitle{SETI Strategies}
\shortauthors{Jason T.\ Wright}

\graphicspath{{./}{figures/}}

\begin{document}

\title{Strategies and Advice for the Search for Extraterrestrial Intelligence}

\correspondingauthor{Jason Wright}
\email{astrowright@gmail.com}

\author[0000-0001-6160-5888]{Jason T.\ Wright}
\affil{\PSETI}
\affil{\PSUAA}
\affil{\PSUCEHW}

\begin{abstract}

As a guide for astronomers new to the field of technosignature search (i.e.\ SETI), I present an overview of some of its observational and theoretical approaches.

 I review some of the various observational search strategies for SETI, focusing not on the variety of technosignatures that have been proposed or which are most likely to be found, but on the underlying philosophies that motivate searches for them.  
 
 I cover passive versus active searches, ambiguous versus dispositive kinds of technosignatures, commensal or archival searches versus dedicated ones, communicative signals versus ``artifacts'', ``active'' versus derelict technologies, searches for beacons versus eavesdropping, and model-based versus anomaly-based searches.  I also attempt to roughly map the landscape of technosignatures by kind and the scale over which they appear. 
 
 I also discuss the importance of setting upper limits in SETI, and offer a heuristic for how to do so in a generic SETI search. I mention and attempt to dispel several misconceptions about the field.
 
 I conclude with some personal observations and recommendations for how to practice SETI, including how to choose good theory projects, how to work with experts and skeptics to improve one's search, and how to plan for success.
\end{abstract}

\keywords{keywords}

\section{Introduction}

\subsection{Context}

The modern era of the Search for Extraterrestrial Intelligence (SETI) began around 1960 with the realization that humanity was capable of producing signals that humanity was capable of detecting at interstellar distances. \citet{cocconi_searching_1959} and \citet{Schwartz61} showed that radio and laser transmissions would very soon meet this standard, and in 1960 Frank Drake began the first SETI Program, Project Ozma, in the radio \citep{OZMA,DrakeEquation}. At around the same time, \citet{Bracewell60} argued that the best strategy would be to look for probes within the solar system, and \citet{dyson60} suggested looking for waste heat from technology orbiting other stars. Since then, a variety of ideas and approaches have been suggested and conducted, and many arguments about the success of SETI have been advanced.

Most of the progress in the field since then has been in radio SETI. In the US, this progress was first at NASA and then later at the SETI Institute after Congressional opprobrium ended federal support for the project \citep{Garber99,Tarter01}. Work at the SETI Institute relied primarily on funds from major philanthropic donors, first from Barney Oliver, who helped found the institute and whose funds supported Project Phoenix, and then later Paul Allen who funded construction of the Allen Telescope Array (ATA).

In academia, most of the work was in the radio at the University of California, Berkeley, which developed the SERENDIP and SETI{@}Home projects, and developed technology for the ATA \citep{Werthimer2001}; and at Ohio State, where the Big Ear was used to search for signals for over 20 years, and where the ``Wow! Signal'' was detected \citep{Wow}. At Harvard, Paul Horowitz led several projects in laser SETI looking for pulsed signals \citep{Horowitz1978,howard04}.

There is a broad sense today that the field is undergoing a resurgence, spurred by several happy developments. One is the discovery of exoplanets, and the determination that the $n_e$ term in the Drake Equation (i.e.\ the average number of Earth-like planets per star) is on the optimistic end of estimates, which significantly increases pessimists' estimates of the number of potential signals there exist to find. Second is the Breakthrough Listen Initiative founded by Yuri Milner \citep{Worden2017} and executed by UC Berkeley, which has greatly increased the amount of searching done, the number of people trained in the field, the visibility of the field, and the opportunities for practitioners to collaborate and contribute to the effort.

Third has been a broadening of the focus of the field in the 2010's beyond the laser and radio projects that could find support in the limited funding environment of the '80's, '90's, and aughts, inspired by the flourishing of the field of astrobiology under NASA's aegis. \citet{tarter06} argued, rightly, that SETI belonged under the same astrobiology ``umbrella'' as other ways to search for life, and coined the term ``technosignatures'' by analogy to ``biosignatures'' to emphasize the parallel and complementary approaches of the fields.  

This rebranding of SETI as the search for technosignatures seems to have helped.  Congressional opposition to searches for technological life seems to have abated and even reversed, and draft language for support for the field in NASA appropriation and authorization bills in Congress has inspired NASA to support the field once again \citep{HoustonTechnosignatures}, especially in directions beyond radio and laser SETI.

This new framing of the problem has led to a perception that the field has changed direction, that old paradigms have been shattered, that the old assumptions and fallacies of the field are finally being discarded in favor of a new, better SETI. But while it is certainly true that there is much new energy, many new ideas and approaches, and more and broader participation in the field than even five years ago, this perception is an exaggeration. A review of the literature and practice of SETI today reveals that SETI is and always has been a broad and dynamic field, and its narrowness and slow progress of the past were primarily a consequence of insufficient and narrow funding, not a lack of imagination or scientific rigor.

In this document, I hope to inspire those that would enter the field to find the approach that resonates with them best, so that they might pursue SETI according to their talents and interests. To this end, I present a perhaps idiosyncratic review the various observational search strategies for SETI, focusing not on surveying all of the proposed technosignatures or opining on which are most likely to be detected, but on the underlying philosophies that motivate searches for them.  I conclude with some personal observations and recommendations for which sorts of projects are ripe for new work (and which are not), and how to practice SETI generally.

\subsection{The Theory and Practice of SETI}

Progress in SETI, as in most of astronomy, proceeds on three complementary fronts: theory, instrumentation, and observation.  

The theory of SETI is dominated by discussions of two foundational concepts: the Drake Equation \citep{DrakeEquation} and the so-called Fermi Paradox \citep[see, e.g.][]{FermiParadox,cirkovic18}.  

Work on the former generally involves attempts to calculate the value or uncertainty on $N$, the number of detectable species in the Galaxy, given what we do and do not know about the prevalence and character of life in the Galaxy. Work on the latter generally involves suggestions for explaining why Earth has not been visited, and what, if anything, this implies about the chances of success for SETI. By analogy, it also includes work on why we have not yet found evidence of technological alien life, such as studies of the Great Filter \citep{Filter} or the ``hard steps'' in the development of intelligence \citep{HardSteps}.

It is important to note that the Drake Equation and the Fermi Paradox make mutually incompatible assumptions about alien behavior: the Fermi Paradox assumes that technological life \textit{will} spread throughout the Galaxy, while the Drake Equation assumes it will stay put \citep[although many authors have worked to blend them, e.g.][]{Prantzos2013}. 

Work on both topics is essential for justifying the endeavor, interpreting null results (and, someday one hopes, detections), and situating SETI within the broader astrobiological and scientific landscape. But such work fundamentally operates from a position of extreme ignorance: we have a sample size of exactly one when it comes to life in the universe, and since that one is humanity and life on Earth, anthropic considerations render extrapolations from even that one point dubious \citep[e.g.][]{Kipping2020a}. 

As well put in an international petition by Carl Sagan and 72 other scientists (including SETI skeptics who argued that SETI would fail): 
\begin{quote}
\textellipsis the only significant test of the existence of extraterrestrial intelligence is an experimental one. No \textit{a priori} arguments on this subject can be compelling or should be used as a substitute for an observational program. \citep{SaganPetition}
\end{quote}

Theory more closely related to the observational practice of SETI---that is, the actual searching for technosignatures---includes discussion of what forms alien technology might take, its scale, its location, and what technosignatures it will produce. From this work we now have many proposed technosignatures and search strategies---many more than have actually been searched for and executed.

Instrumentation in the field has progressed swiftly, alongside instrumentation across astronomy and planetary science. While searches for many technosignatures can be made with the kinds of data sets collected for generic astronomy, many technosignatures, especially those involving communicative signals, benefit from dedicated hardware.

Searches in the radio have expanded with the advent of broadband receivers and backends \citep[e.g.\ the Breakthrough Listen backend at the 100m Green Bank telescope][]{MacMahon18}, and multi-dish arrays that can greatly mitigate radio frequency interference \citep[e.g.\ the Allen Telescope Array][]{Tarter2020,Harp2005}. These have greatly reduced the need to guess at ``magic frequencies'' and continue to grow rapidly in power as computational capabilities increase.

Work in the optical and near-infrared has seen orders-of-magnitude improvement in the past few years in the capacity to detect time-compressed signals. The state of the art today is PanoSETI \citep{PANOSETI}, which will be sensitive to nanosecond-scale pulses over half of the entire sky simultaneously. 

In this work I will focus on the third front, actual searches, as it has been enabled by the other two (although I will return to the theory aspects of SETI in Section~\ref{sec:recommendations}).  My aim is to provide categories to organize the field and provide some structure within which search strategies can be compared and contrasted.  

\subsection{The Merit of Different Approaches}

\citet{Sheikh2020} provided an excellent organization of the field in terms of nine axes of merit commonly used to justify searches for different kinds of technosignatures.  Four of these axes were purely practical and a function of \textit{us} and \textit{our} technology: cost, observing capability, ancillary benefits, and detectability or sensitivity. Obviously, searches that score well along these axes are easier to pursue, and thus more meritorious.

The other five axes are functions of the technology we seek to find: duration (essentially Drake's $L$), ambiguity, extrapolation from human technology, inevitability, and information content.  Technosignatures that score highly on these axes are more ubiquitous, easier to discover, and more infomative, and therefore more meritorious.

My purpose in this work is not to use these axes to rank or compare the merit of different technosignatures, except incidentally. My goal, rather, is to describe an objective set of criteria by which to organize technosignature search strategies and differentiate their underlying philosophies, regardless of whether they might be a good idea. Nonetheless, my analysis will examine many of \citeauthor{Sheikh2020}'s axes because they capture well different approaches to the problem, and so some judgement of merit is unavoidable. 

In section~\ref{sec:recommendations} I will close with recommendations, but not regarding which technosignatures one should look for. In that regard, I will simply observe that we do not know which methods are most likely to succeed, so we should pursue many and avoid unnecessary subjective assessment of their merit when \citeauthor{Sheikh2020}'s axes do not provide an objective and unambiguous ranking. 

\section{Communicative Signals Versus Artifacts}

\subsection{Communicative Signals}

The first SETI searches were for communicative signals, mostly in the radio and later in the infrared and optical motivated largely by the necessity of looking for beacons, partly by the desire to look for what were Earth's loudest technosignatures at the time, and partly by hopes or expectations of contact and communication with alien life.

Communicative signals would be designed to maintain their fidelity in the presence of natural sources of noise, and so must have some property that allows one to distinguish the photons (or other carriers) from those emitted by a natural source. This is especially important if the signal is emitted from, say, a small device on a planet orbiting a star, because the star is constantly emitting copious photons across a huge range of wavelengths, and the small device cannot hope to compete without some sort of distinguishing feature.  
i
One option is to use a wavelength where most stars are dark, such as radio waves. In the radio, choosing photons of higher energies  leads to difficulties as the information that survives the 1/r$^2$ law per erg gets lower, and moving to lower energies introduces the problem of background photons from electrons in the interstellar medium. The balance of these two issues historically led to the development of the ``Water Hole'' \citep{WaterHole} as an optimal frequency in which to transmit and search, but the case is somewhat overstated: more recent work has found benefits to using a wide range of wavelengths, especially optical or infrared laser light (see \citet{Hippke18_Benchmarking} for a discussion of the merits of other frequencies and carriers.)

The most commonly proposed ways to make a signal stand out against noise are by compression in the time or frequency domains (indeed, this is how much communication on Earth proceeds). This works because the photons from background or noise sources are spread out over a huge range of frequencies and times, and so packing even a small number of photons into a very narrow range of either can make them stand out. Transmitting and receiving photons at either high time or frequency resolution thus dramatically lowers the energy requirements for outshining a star or other source of noise.

Operating at high time or frequency resolution has the added benefit that such a signal is unambiguously technological, because natural sources of photons have finite size, and are often gases with both thermal and internal bulk velocities (e.g., from turbulent or rotational motions).   The finite size of emitting regions puts a minimum on the duration for any signal or change in flux set by the light crossing time of the region.\footnote{A caveat here is that  relativistic beaming, as from a jet of material pointed towards earth, can compress a signal temporally and so evade this minimum. Such sources are typically not going to be confused for communicative signals, however.} Since a natural source must be rather large to emit enough photons to be detectable at interstellar distances, its signals must be spread out in time. Thus, any burst of photons arriving on timescales of extremely small fractions of a second must have been engineered to do so. 

Compression in frequency is the opposite extreme (in a Fourier sense): the thermal and other motions of a gas mean that any photons it emits will inherit a range of Doppler shifts, putting a lower limit on the spectral width of any emitted light, even from, for instance, masers, which are very narrow in their rest frame. Any line with a width less than the thermal velocity of any gas that could emit the photons must be technological.

Searches for such communications thus score highly on the axes of merit in that they can be unambiguous and are potentially information rich.

One complication is that modern human communication schemes employ a wide variety of modulation and compression schemes depending on context. We might then expect that alien transmissions would similarly be complex, and defy simple description in terms of time or frequency compression. This concern is somewhat mitigated by the context of interstellar transmissions, which must account for huge signal strength loss due to the $1/r^2$ law, ubiquitous background photons, and significant Doppler drifts, few of which are concerns for Earth-bound transmission.

Nonetheless, we mustn't assume that we have identified the optimum solution to this problem, or that alien technology would use it, especially if we wish to eavesdrop and not only look for beacons. Searches for communicative transmissions can thus employ a combination of model-based searches based on our expectations, as well as more general searches for other kinds of signals.

\subsection{Artifacts}

\citet{WrightAdHoc} define ``artifact SETI'' as ``the search for physical manifestations of technology, exclusive of communicative transmissions.''  This is then a broad category, being virtually any technosignature that is not a communicative signal.

The most literal form of artifact SETI is the search for physical artifacts in the Solar System, and this is indeed how \citet{Freitas_SETA} introduced the (now deprecated) term SETA (The Search for Extraterrestrial Artifacts). This part of the search is motivated by the same assumptions behind the Fermi Paradox: that if technological life exists, it will be spacefaring and eventually explore the entire Galaxy, including the Solar System. We should therefore look for evidence of such exploration, either in the form of free-floating probes or artifacts on or under the surfaces of planets.  

At the other extreme, \citet{cirkovic2006} and \citet{Bradbury11} describe ``Dysonian SETI'' as the search for megastructures or similar mega-engineering projects in the Galaxy and beyond, such as Dyson spheres.  Such artifacts might be detected via their waste heat, for instance, as \citet{dyson60} originally suggested.

Physical artifacts could generate a variety of other technosignatures, including artificial pollutants in atmospheres, high energy particles, quickly moving objects or stars, or bursts of radiation designed for propulsion. Indeed, since we presume any technosignature will involve a physical underlying technology, it can be difficult to circumscribe the limits of what it means to search for ``artifacts''. 

The primary difference between communication and artifact SETI comes then from the rationale for the search.  In communication SETI, the target of the search is the signal itself, and the nature of the transmitter is incidental. The technosignature in this case is the target of the search. In artifact SETI, the target of our search is the underlying technology itself, and the technosignature merely the means to that end.

\subsection{Extant Versus Derelict Technology}

Technology might remain detectable long after it has fallen into disrepair.  For instance, the stability of a Dyson sphere has not been thoroughly studied \citep{Wright2020_Dyson} but it might last as a generator of waste heat long after its creators are gone. Artifacts in the Solar System might be active, producing communicative signals, generating heat, maintaining attitude, and maneuvering; or they might be long dead, subject to collisions with asteroids or meteorites, or lying beneath the regolith or sands of a terrestrial body.

We do not know the limits of technology's ability to self-maintain. If we detect a communicative signal, it will likely have been transmitting for a long time (Drake's $L$ must be large in order for $N>1$). But this does not necessarily imply that its creators have been around that long; it may have simply been built to last.  Indeed, technology itself might even be able to adapt, replicate and grow much like life does, completely blurring the line between bio- and technosignatures.

\subsection{Scales and Kinds of Artifact Technosignature}

One can, at least roughly, map the artifact technosignature landscape by the form of the technosignature and the scale of the technology itself.  In general, the more distant a technology is, the grander its scale need be for us to detect it. The first work to quantify technosignatures in this direction was the scale of \citet{kardashev64}, who based it on energy consumption, and this work has been generalized by many, including \citet{Zubrin00}, to refer simply to a physical scale.

In the following, I will offer some examples of technologies and some characteristic references for them, but this list is not meant to be comprehensive, nor is it meant to reflect the most plausible or detectable technosignature in each category; they are merely illustrative of how one might categorize the range of the field.

At the smallest scales/distances we would have searches within the Solar System, for instance for physical artifacts such as probes, but also for things like city lights \citep{Loeb11} or reflections from solar collectors \citep{Dyson2003}. We might also look for chemical evidence of past technology, perhaps recorded in the Martian ice caps or in planetary atmospheres. If there is active technology using energy, it could be detected by its thermal emission. Such forms of SETI have many synergies with work in planetary science.

On larger scales we could look for technosignatures associated with exoplanets, such as their satellite belts \citep{Socas-Navarro18,Sallmen19}, structures at their Lagrange points \citep{Korpela15,gaidos17,Clara}, or even directly imaged on their surfaces (for instance with the Solar gravitational lens).  The planets themselves might have atmospheric technosignatures \citep{Lin14,Kopparapu2021} akin to NO$_\mathrm{x}$s or chlorofluorocarbons. \citet{kuhn2015global,Kuhn18,Berdyugina19} have explored how one might identify rotationally modulated hotspots, for instance due to cities. Much work in this direction might occur in concern with other work to study exoplanets in general. These technosignatures roughly correspond to \citeauthor{kardashev64}'s Type {\sc i} species.

Circumstellar material might be on a scale even larger than planets, such as a Dyson sphere \citep{Wright2020_Dyson}, either around an ordinary star or potentially around other objects as well \citep{Semiz15,Osmanov16,Imara18} Such material can be detected via its waste heat, or by the starlight it blocks \citep{GHAT4}. Other suggestions for stellar-scale technosignatures include stellar atmospheric pollution \citep{Whitmire80} or ``stellar engines'' that move the stars themselves \citep{badescu2000,Svoronos2020}. These technosignatures roughly correspond to \citeauthor{kardashev64}'s Type {\sc ii} species. 

Finally, we can follow the original spirit of the Fermi Paradox and imagine pan-galactic technologies, considering \citeauthor{kardashev64}'s Type {\sc iii} species. Indeed, as \citet{GHAT1} pointed out, if the SETI pessimists are correct and the lack of any obvious galaxy-scale engineering proves that we are alone in the Milky Way, then unless we are alone in the \textit{Universe, other} galaxies should show signs of such engineering. We might search for waste heat across whole Galaxies \citep{GHAT2,GHAT3}, stellar population management (for instance, to suppress cataclysmic events such as gamma ray bursts or supernovae), or other forms of galactic engineering. Certain very luminous sources might even be visible at such distances \citep{Lingam17_FRB}.

\begin{figure}
    \centering
    \includegraphics[width=\textwidth]{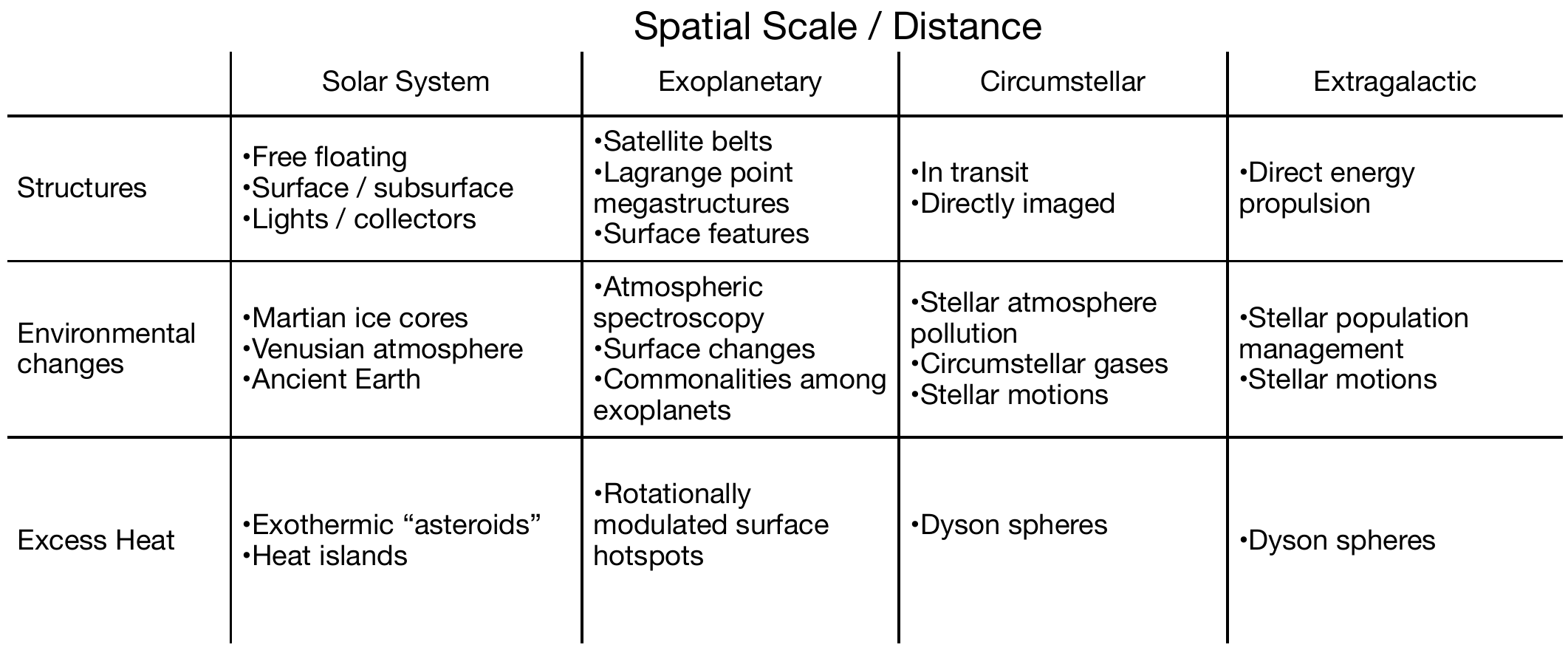}
    \caption{Some potential technologies and artifact technosignatures grouped according to their scale and their kind.}
    \label{fig:chart}
\end{figure}

We might then group these various artifact technosignatures into three broad camps: detecting artifacts themselves via the light they emit or block, detecting the changes they have on their environment, and via the excess heat they generate from energy use.  This grouping is expressed in Figure~\ref{fig:chart}.

\section{Searches for Ambiguous vs.\ Dispositive Technosignatures}

As with biosignatures, the issue of how one can know that one has succeeded in detecting life after detecting what appears to be a technosignature is complex.

A great virtue of searches for communicative signals is their prospect for being dispositive: any signal from beyond the Earth that is sufficiently compressed in time or frequency must be artificial. Such a detection would scale the entire ``Ladder of Life Detection'' at once \citep{Ladder}, showing that complex life that has produced detectable technology exists elsewhere in the universe.  

Such a detection also offers the prospect of more detailed study than an ambiguous detection. In the most extreme case, the signal may contain decipherable information. Such a signal could then include anything as grand as plans for powerful technology or the ``Encyclopedia Galactica,'' or as intimate as the instant thoughts of a single being at the controls of the transmitter.  

But even absent such content (either because it is undecipherable, or because it is absent), the mere presence and form of an unambiguous signal can be studied in the same way as biosignatures to deduce the properties of the technology itself, and the nature of its environment. Knowing its form would also motivate additional searches for similar phenomena elsewhere (either because the technology could be from a spacefaring species that spread it far and wide, or because it represents an engineering optimum for a particular problem that is appreciated by many species).

Of course, there is no guarantee that unambiguous technosignatures are detectable, and there may be many more and more easily discovered ambiguous technosignatures.  Dyson spheres, for instance, will share many (but not all) of the characteristics of natural circumstellar material \citep{GHAT2,GHAT4}, and the spectral signatures of atmospheric technosignatures like CFCs might be similar to that of combinations of purely natural gases.

Searches for ambiguous technosignatures such as these will have a longer road to hoe to prove the discovery of alien life, one that will largely proceed as natural explanations are slowly ruled out \citep{GHAT1}.  Still, it can serve to generate candidate techosignatures for follow up by more dispositive forms of SETI.

Searches for unambiguous technosignatures may be less dispositive than they seem at first glance, however. When a novel anomaly is first discovered it may seem to be so strange at first as to be inexplicable except as technology. Theorists are a clever bunch, however, and what seems inexplicable at first may soon have more natural explanations than data.  The discovery of the interstellar comet 1I/`Oumuamua, for instance, at first generated much puzzlement and, naturally, some speculation that it could be artificial \citep[probably fueled by its similarity to \textit{Rama}, the spacecraft from an Arthur C.\ Clarke novel][]{Bialy18}.  Rather quickly, however, planetary scientists were able to show that it fit within the bounds of expectations for natural objects \citep{ISSA_Oumuamua}.

A third approach is to search for extremely ambiguous technosignatures, by pursuing a SETI interpretation of objects that \textit{do} seem to have satisfactory natural explanation. \citet{Vidal2016}, \citet{Vidal2019}, and \citet{Haliki2019} have explored the possibility that certain kinds of binary stars or pulsars might be technological (or, even, living things themselves).  This approach can have the disadvantage that not only might it be challenging to test the hypothesis, but the motivation for the hypothesis in the first place may be unpersuasive to many. On the other hand, one would hardly want to miss a technosignature just because it had been misinterpreted as natural, so such work still has some merit, provided it is well grounded. 

\section{Commensal / Archival versus Dedicated Searches}

Because SETI has historically been funding-starved, it has often had to ``piggyback'' on other projects, as, for instance, with the SERENDIP project at Arecibo, which observed simultaneously with other instruments \citep{SERENDIP}.  Such projects often sacrifice being able to point a telescope in exchange for having access to its focal plane at all times.  

Such work has historically been called ``parasitic,'' however this term is now discouraged because it implies that the observatory (or the rest of astronomy) suffers from such projects. A better term borrowed from biology, and more commonly used today, is ``commensal'' implying that SETI benefits from such work while doing, at worst, no harm.  Indeed, even this understates the case: having SETI work for ``free'' at an observatory often expands its science case, its publication record, and its public profile, and so the truth is something a bit closer to ``symbiotic.''

Closely related is archival work, which makes use of data collected for other purposes. \citet{Albrecht1988} and \citet{Djorgovski00} were early proponents of big-data and algorithm-based approaches in sky surveys, a strategy followed by, for instance, \citet{carrigan09a} and \citet{GHAT3} for Dyson spheres, and the VASCO project for disappearing stars \citep{VASCO}. More targeted data is also useful, as with the \citet{Tellis15} and \citet{Tellis17} searches for continuous wave lasers near nearby stars and potential Habitable Zone exoplanet host stars.

A dedicated search has the advantage that it can be designed as a careful survey, and so more easily focus its efforts on likely targets, and more easily calculate upper limits.  Such work can also make use of dedicated hardware that is more sensitive to the technosignature being searched for.

Dedicated SETI searches using competitive time on public telescopes can be challenging because it has not happened much in the past, so such work can seem foreign to many time allocation committees.  This may change if such proposals become more common, and such work begins to have a positive effect on observatories' paper publication rates.

\section{Model-based versus Anomaly-based Searches}

\subsection{Model-based searching}

The most straightforward way to find something is to have a model for what it looks like and look for that. This approach has the disadvantage that it requires ``xenopsychology'' or ``xenosociology'': we must imagine why and/or how the engineers of the technology built it in order to model it, and in order to justify its existence to motivate the search. But a great strength of this approach is that it straightforwardly allows for upper limits on the technosignature to be calculated.

\begin{figure}
    \centering
    \includegraphics{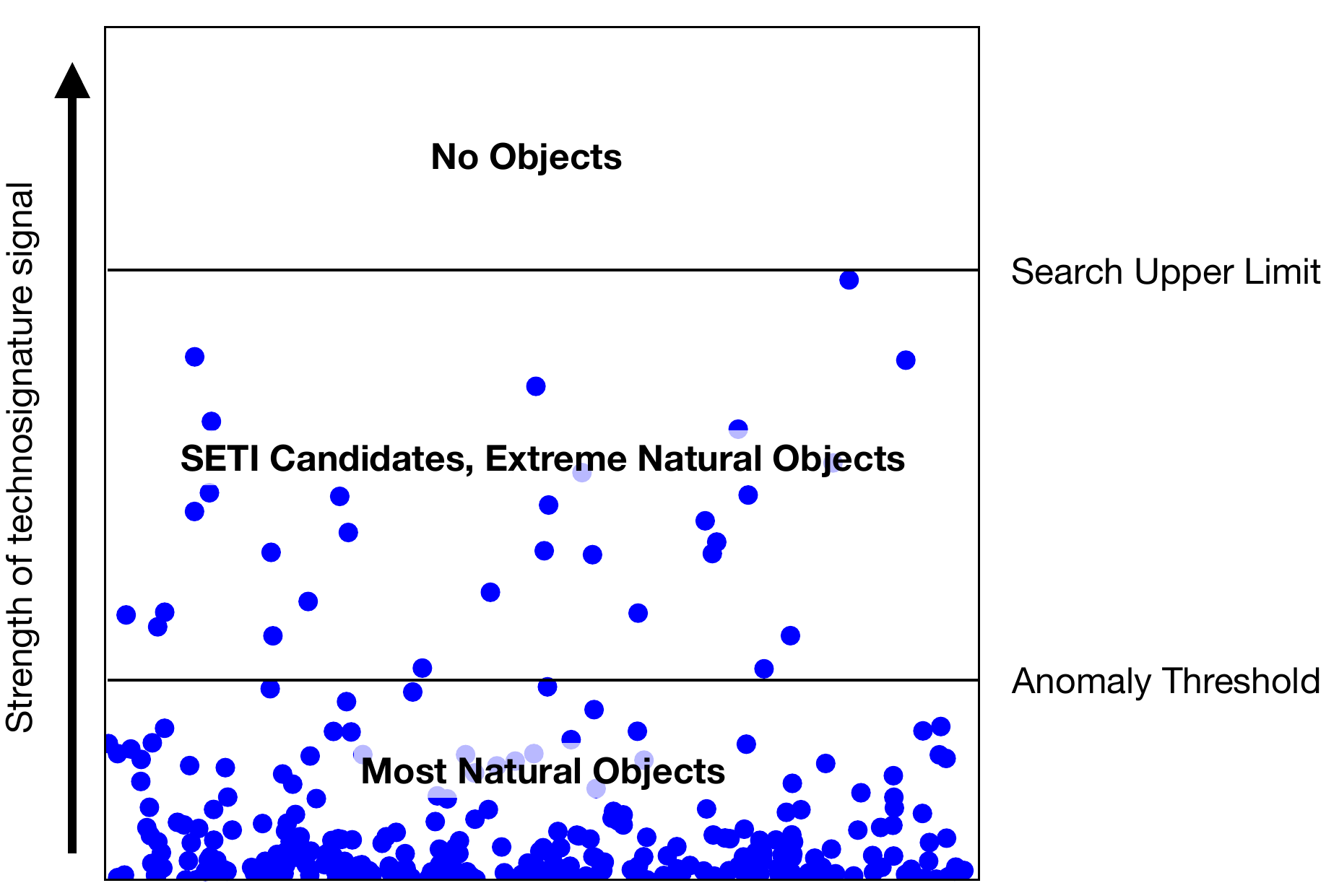}
    \caption{Heuristic illustration of how upper limits can be set in SETI for a model-based search with natural confounders, such as a search for Dyson spheres. One makes measurements of many objects provisionally interpreting all signals as potential technosignatures, and ranks them by their strength. Above some threshold no objects exist, and this provides a first, weak upper limit on the prevalence of that technosignature. An anomaly threshold is set by either the sensitivity of one's search, or at a higher level that provides a manageable number of candidates. One then explores the candidates between the threshold and the upper limit with complementary measurements that can be expected to distinguish natural from artificial causes for the signal, or at least confirm that the object is a member of a known, natural class and that the hypothesis of artifice is not warranted. As the most extreme natural objects become identified as such, one's upper limit drops.  More thorough searches can then improve on this limit through better sensitivity or better follow-up. Ancillary science can then be performed on the extreme natural objects, which are likely to be inherently interesting simply because they are extreme.}
    \label{fig:upperlimits}
\end{figure}

Presuming that one can express the strength of such a technosignature along a single axis, one can schematically illustrate that strength among all objects in a catalog, for instance from an all-sky survey, as in Figure~\ref{fig:upperlimits}.

Even before any analysis of individual objects is undertaken, the most extreme object in the catalog represents a first upper limit on the technosignature.  Of course, this upper limit may be extremely weak, but that is the nature of first upper limits.  As a referee complained on an early, weak upper limit on neutrinos in a manuscript by \cite{Davis1955}:

\begin{quote}
Any experiment such as this, which does not have the requisite sensitivity, really has no bearing on the question of the existence of neutrinos. To illustrate my point, one would not write a scientific paper describing an experiment in which an experimenter stood on a mountain and reached for the moon, and concluded that the moon was more than eight feet from the top of the mountain. \citep[][p. 245]{neutrinos}
\end{quote}

\citeauthor{Davis1955} would go on to win the Nobel Prize for the discovery of neutrinos, and the ultimate success of that project shows that the referee had undervalued the of setting weak limits as part of the process of discovery.

Likewise, the first upper limits on technosignatures, especially those that have not been explored before, may at first seem uselessly weak. We should welcome such upper limits as important first steps, and as challenges to others to improve upon them.

For instance, consider the hypothesis that there are alien artifacts on Mars, which once had significant scientific currency. Indeed, the idea that there were living cities and global agriculture on Mars was entertained seriously by at least some astronomers well into the twentieth century, and not fully dispelled until the Mariner missions revealed a surface devoid of any obvious signs of such technology \citep{Sagan1976}. In hindsight, the idea of cities on Mars seems laughable, like reaching the moon from a mountaintop, but that is the nature of hindsight.

Today, we can be sure that there are no artifacts larger than some size on Mars, but a proper quantification of that limiting size has never been performed. In order to generate such a quantification, specific observable qualities of artifacts will need to be proposed, modeled, and searched for in maps of Mars. When the first upper limits on Martian artifacts are submitted for publication, they may be quite weak because the assumptions one must make to do a search may be quite strong, and one can imagine a referee's response similar to Davis's. But one must start somewhere.

Once this upper limit has been established, it can be improved upon by examining the most extreme objects just below this limit.  In general, such objects will be rare and interesting exactly because they are extreme (whatever their nature). As their nature becomes understood, they can be removed as SETI candidates from the list (or the model that measures technosignature strength can be modified to accommodate them as a source of ``background'') and the upper limit can be improved.

Of course, upper limits are only useful to others if they are published. For many years, very few SETI searches published quantitative upper limits, and these must therefore be inferred and estimated based on the known properties of the instrument and the published properties target lists. Fortunately, the community has become much better about publishing its results, and several schemes for announcing upper limits have been developed \citep{howard04,Enriquez17,Haystack}. Technosearch, a resource developed by Jill Tarter and others, tracks both inferred and explicit upper limits in SETI can be found at \href{http://technosearch.seti.org}{technosearch.seti.org}.

\subsection{Anomaly-based searching: ``generalized SETI''}

Model-based searches have the strong disadvantage that they require us to guess or deduce the form of alien technology---if not technology-as-we-know-it then technology-as-we-can-imagine-it. It is common to the point of clich\'e in such discussions to invoke Clarke's Third Law, that ``any sufficiently advanced technology is indistinguishable from magic'',\footnote{And Gehm's corollary: ``Any
technology distinguishable from magic is insufficiently advanced.''} meaning that we should not model alien technology on things we understand, but instead look for apparent violations of the laws of physics, or at least physics that does not account for technology.  

An alternative to model-based searching, therefore, is to assume that technosignatures are rare but may be evident as highly unnatural phenomena, and look for outliers along any arbitrary dimension in astronomical catalogs. This approach was articulated well by \citet{Djorgovski00}, who referred to it as ``generalized SETI.''

Such searches might seem at first to stretch the very definition of SETI: if one is merely looking for that which one does not understand, then one is not explicitly searching for alien technology, but is merely looking for the unusual, not necessarily the technological. But this approach is well-motivated by the fact that any persuasive detection of alien technology will necessarily have to be so different from any plausible natural source, that only technological origins make sense. Any such signal should, to a sufficiently sensitive anomaly-detection algorithm,
stand out quite cleanly.

Indeed, one can then proceed in much the same way as described in the last subsection, examining the strongest outliers among many axes to find both extreme natural objects and good SETI candidates.  The primary difficulty is that it may be hard or impossible to model exactly what the ``strength'' of a technosignature parameterizes.  But because it presumes little or nothing about the nature of alien technology, this generalized approach may be our strongest tool in avoiding human biases in what alien technology will be like.

\section{Searching for Beacons Versus Eavesdropping}

Many of the earliest searches for technosignatures were searches for ``beacons''---signals intended to be discovered, perhaps specifically targeted at Earth itself.  Such technosignatures do well on many of the \citeauthor{Sheikh2020} axes of merit because they should be strong, obvious, long-lived, and unambiguous, by design. Indeed, the only axis they might do poorly on is that of contrivance/inevitability.

Such searches are contrasted with ``eavesdropping,'' or the search for technosignatures not intended to broadcast one's existence.  Such technosignatures form a broad category, from deliberate transmissions not intended for us (for instance leaked radio transmissions), to consequences of industry (for instance thermal radiation from waste heat or atmospheric pollutants), to relics and artifacts (such as probes or artifacts in the solar system.)

Searches for beacons would seem to be much more likely to succeed than other kinds of searches \textit{if} such signals exist. The primary difficulty in detecting them will likely lie not in achieving the requisite sensitivity (since the signals should, by design, be quite strong), but in selecting the correct form, location, and time of the transmission.

Determining these parameters is a well known concept in game theory, and is accomplished via ``Schelling points,'' i.e.\ equilibria in a non-communicative, cooperative game such as mutual search \citep{schelling1960strategy,Wright2020_Planck}.  Such an approach motivates the selection of every aspect of a search strategy: the wavelength range, the kind of detector or telescope, when and where to observe, and what kind of modulation to search for.  

There is a misperception that early SETI programs, especially in radio SETI, were motivated entirely on the optimistic assumption that there exist beacons to be found. In fact, the pioneers in the field recognized that success might come not from the detection of a bright, narrowband signal meant for us, but from indistinct ``chatter'' of leaked communications or other sources of radio transmissions. \citet{DrakeEquation}, for instance, described a method for identifying sub-threshold signals in radio data via cross-correlation of signals taken at different times, and \citet{Dyson66} described looking for the ``stray electro-magnetic fields and radio noise produced by large-scale electrical operations.''

The focus on beacons at the time was largely driven by the necessity born of the infancy of the field. First, beacons are the ``low hanging fruit,'' being the easiest (and, at the time, practically only) signals one could search for. One naturally searches for them first. Second, the limitations of instrument bandwidth and channelization at the time meant that practitioners had to guess at ``magic frequencies'' \citep{Tarter1980}, using game theoretical and astrophysical considerations such as the Water Hole \citep{Oliver79}.

It is important to recognize that searches for many or most technosignatures are sensitive to \textit{both} beacons and other technosignatures, such as leaked emission. For instance, \citet{Arnold05}'s suggestion of looking for transiting megastrutures in \textit{Kepler} data works fine as a perfectly general suggestion, but his specific application was to using such megastrutures as beacons. The Breakthrough Listen radio search is sensitive to not just radio beacons, but also signals not meant for us that we might intercept, or the radio emission of non-communicative radiation such as that used for directed energy propulsion systems \citep{Lingam17_FRB}. The strategy to search for beacons thus need not be \textit{exclusive} to such technosignatures.

\section{Passive versus Active SETI (METI)}

One of the more salient and contentious differences in SETI philosophy is whether one should passively search for technosignatures, or solicit a response.  This is the difference between ``passive'' SETI (usually just called ``SETI'') and ``active'' SETI (usually called ``METI'' \citep[for ``Messaging Extraterrestrial Intelligence''][]{Zaitsev2006_METI} by tradition and for syntactic clarity \citet{WrightAdHoc}.)

Passive versus active SETI has a close analogy to active versus passive sonar: purely passive detection requires only that one ``listen,'' while active detection involves some sort of transmission that generates a detectable signal in return. In sonar, that transmission can be a ``ping'' which is heard in reflection, while in active SETI the transmission can take many forms, all of them intended to provoke or invite some sort of detectable response.

METI has a variety of forms. In the sense that it describes the construction and delivery of messages intended for extraterrestrial species, it includes the Arecibo Message \citep{AreciboMessage}, the Pioneer Plaque \citep{PioneerPlaque} and the Voyager Golden Record \citep{VoyagerRecord}. The primary value of these exercises has been the awareness they have raised about SETI, and the framework they provide for discussions about interstellar contact and communication. In this sense, they have been extremely successful, having become staples of college courses and popular discussions of SETI ever since. But as actual signals to other species, they have been essentially useless: the probes are too small, too inconspicuous, and traveling too slowly to have any reasonable chance of being discovered; and the Arecibo message was too weak, too brief, too targeted (and, frankly, too obscure) to provide any practical chance of generating a response.\footnote{This was by design. ``As the choice of frequency, duration of message, and distance of the target clearly shows, the Arecibo message is very unlikely to produce interstellar discourse in the foreseeable future. Rather, it was intended as a concrete demonstration that terrestrial radio astronomy has now reached a level of advance entirely adequate for interstellar radio communication over immense distances. More extensive attempts at the transmission of radio messages from the Earth to extraterrestrial civilizations should be made only after international scientific consultations, and recommended by the Soviet-American conference on communication with extraterrestrial intelligence.'' \citep{AreciboMessage}}

Indeed, the finite speed of light makes the utility of any signal intended for interstellar audiences dubious, and in fact ethically questionable: if the ``ping time'' of such a signal is of order centuries, then sending such a signal is writing a check that only our distant descendants can cash. 

METI's primary value is made clear by what \citet{Zaitsev2006_Paradox} calls the ``SETI Paradox'': contact via searches for deliberate beacons can only succeed if technological species transmit signals that can be found. If we wish to make contact, we must therefore either participate in that two-way communication, or take a purely passive role and hope that some other species take an active one.

Unlike with sonar, however, the boundaries between active and passive SETI are not as clear as they might seem. \citet{Cortellesi2020} describes a ``continuum of astrobiological signaling'' in which Earth life's presence is made clear through a variety of detectable signatures, including biosignatures, leaked radio transmissions, and deliberate signals as part of dedicated METI projects.  

At one extreme of METI is the attempt to make humanity more salient by producing the sorts of beacons that we search for, thus directly resolving the SETI Paradox. This has been promoted both as an important activity in its own right, and as a useful way to engage the public in SETI regardless of its success in helping find alien life \citep{Vakoch2011,Penny2012}.

Such activities, to be successful, have two requirements: that the signals sent be more obvious and detectable than our leaked emissions and signatures, and that we be prepared to receive any potential response.  Neither quality has ever been achieved by any METI project: all deliberate transmissions to date have been extremely brief and narrowly targeted, and not of significantly more power than, say, a typical interplanetary radar transmission; and there is no broad effort to monitor the targets of such transmissions for similar responses after one ping time \citep[although, to be fair, the first responses that could be expected are not due to arrive for many years][]{Vakoch2011}.

This form of METI is somewhat controversial, because of fears by some that the Galaxy is a hostile environment for life, or at least technological life, and that if nearby species became aware of humanity's existence they would come here and harm us. This argument has been made most forcefully by \citet{Gertz_METI, Gertz18}, who advocates a total prohibition on the practice, while others such \citet{Billingham2014} have advocated for a moratorium until we understand the problem more, and others such as \citet{Baum11} have simply encouraged that messages be composed with great care so as not to provoke a hostile response. 

These admonishments have generally had the effect of limiting METI activities to those who feel strongly enough about it to engage in the debate and risk condemnation. \citet{Cortellesi2020} dubs the two camps in this spirited and often angry debate ``metiists'' and ``passivists.'' 

Viewed in this light, METI within the solar system is the most extreme end case. \citet{Bracewell60} recommended pinging suspicious objects in the solar system as a way to determine if they are probes, in order to activate them to deliver their message to Earth. This form of METI is the most provocative because it avoids all of the difficulties of other forms of METI, being highly targeted and powerful compared to our leaked transmissions, and having the potential for generating direct contact almost immediately. Indeed, if such an experiment were successful it could achieve the highest possible score on the ``Rio 2.0'' scale of SETI detection significance \citep{Forgan2019}.

An intermediate form of METI is exemplified by the proposal of \citet{demagalhaes2016} that we embed within our existing transmissions an invitation for any eavesdropping species to contact us.  The reasoning here is inspired by the Zoo Hypothesis of \citet{Ball1973}, who posited that the Earth and its inhabitants are well known to extraterrestrial species, but deliberately kept uncontacted because it is part of a nature preserve or zoo.\footnote{This concept is familiar in science fiction, for instance as the motivation for the Prime Directive in the \textit{Star Trek} universe.}  If this is the case, then inviting contact may be successful. The rationale for embedding this invitation in our regular transmissions is that it does not make our presence in the universe more obvious, in a concession to the passivists.

The metiist position is generally that METI is an important complement to passive SETI, and that fears of harm stem from unrealistic assumptions about alien behavior based on unscientific and popular imaginings \citep{Visions}, and a misapprehension of the degree to which life on Earth is \textit{already} detectable because of our biosignatures. \citet{denning2010} discusses the degree to which the METI debate is essentially a new facade on old questions of how to manage a commons, and not really one for which any technical solution exists. Indeed, many authors \citep{Penny2012,Shostak2013,Haqq2013,Korbitz2014} have concluded that the uncertainties involved are simply too large to make any recommendations one way or another.

Finally, SETI practitioners should bear in mind, as \citet{Penny2012} notes, that the success of even purely passive SETI is not necessarily ``safe.'' We do not know what the consequences of a positive detection on humanity will be, but they may not be small, especially for a detection within the solar system. We should not let our expectations for the varied and potentially strong reactions to SETI's success be set by science fiction's and popular culture's depictions of such events any more than we should let those sources set our expectations for what alien life will be like when we find it. Here, social scientists have an important role to play in the practice of SETI \citep[e.g.][]{billingham1999}.

\section{Recommendations}\label{sec:recommendations}

So, given all of these largely orthogonal philosophies of SETI, which are the most promising? Which are the most practical?

I will decline to recommend which \textit{kinds} of technosignatures one should search for---I have my personal opinions and priors, but they are almost certainly as misguided as anyone else's.  Rather, I will give general recommendations on features that any SETI project should follow.

\subsection{Read the Literature}

Because there is no formal SETI curriculum, it is difficult to find and appreciate the substantial body of work that has been done on any given topic within SETI. A good place to start is the SETI bibliography at the NASA Astrophysics Data System described by \citet{Reyes2019} and \citet{Lafond2021}. 

A look at this bibliography shows it contains over 2,000 entries, only half of which have ever been cited and only 200 of which have been cited more than 10 times. The $h$-index of the field is in the mid 40's, and five of the papers contributing to that number only mention extraterrestrial intelligence tangentially or tacitly. One gets the impression that many authors do more writing than reading.

My own experience is that when I have an idea I have never seen in the literature I can almost invariably find an old paper on it with zero or single digit citations, often in the {\it Journal of the British Interplanetary Society} or in some piece of gray literature. Indeed, many of my supposedly novel ideas have turned out to have been developed and published independently multiple times. For the field to grow, authors need to extend prior work, instead of (inadvertently) finding new ways to repeat old arguments.

This is not to say that the body of SETI literature is characterized by thousands of mostly unread papers of high quality: the corpus is highly heterogeneous, and many of the papers in it are obscure for good reason. But even many of the lesser papers contain valuable nuggets, and if SETI is to be a true discipline, its practitioners need to build on the foundation of what has come before, and this begins with the elbow-grease work of literature searches and citation trees, and continues with proper acknowledgement of the priority of ideas.

Such work will help authors appreciate the variety and nuance of the work to date, and prevent one from attacking strawmen. Indeed, I have also found that many common challenges to ideas or assumptions commonly ascribed to ``traditional SETI" are in fact as old as the field itself, and were often shared (or at least articulated) by the pioneers of the field. 

\subsection{Choose Theory Projects Carefully}

Above I have focused on the observational side of SETI, perhaps reflecting my training as an observational astronomer. To a theoretical astrophysicist or astrobiologist looking to work on SETI projects, I recommend those that ``stay close to the data,''\footnote{When I learned this admonition while being trained as an observer in graduate school, it was attributed to my academic ``grand-adviser'' George Herbig.} for instance by using hypotheses about alien behavior to generate new search strategies, determine the detectability and likelihood of success of various proposed technosignatures, or by building frameworks for the interpretation of candidates and null results.

I think it is fair to write that it is \textit{especially} important in a data-starved field like SETI, which has very little to go on, that its theory work stay grounded in observations. In the absence of data, it can be easy to develop a large corpus of work based on assumptions that may one day turn out to be false. While such work can still be worthwhile---there may be spinoff applications, or that work might be part of a broader theory landscape that ultimately lands on the correct way forward---the return on investment is low, which is an important consideration in a field as funding-starved as SETI.

\subsubsection{The Fermi Paradox}

Theoretical work on the Fermi Paradox (and related concepts, such as the Great Filter) is important, but I fear we have approached a point of strongly diminishing returns that will persist until a detection is made. \citet{denning_impossible_2013} writes of the ``now-elaborate and extensive discourse concerning the Fermi Paradox" as being ``quite literally, a substantial body of analysis about nothing, which is now evolving into metaanalysis of nothing.'' She continues, 
``I would not suggest that these intellectual projects are without value, but one can legitimately ask what exactly that value is, and what the discussion is now really about\ldots'' She further writes of early work on this problem that 
\begin{quote}
Thinking about that future [of contact with ETI] was itself an act of hope. Perhaps it still is. But I want to suggest something else here: that the best way to take that legacy forward is not to keep asking the same questions and elaborating on answers, the contours of which have long been established, and the details of which cannot be filled in until and unless a detection is confirmed. Perhaps this work is nearly done\ldots
\end{quote} 

One common trap that has been remarked upon often elsewhere is to attempt to ``solve,'' ``dissolve,'' or ``resolve'' the Fermi Paradox via what \citet{hart75} calls ``sociological explanations," i.e.\ proposing ironclad laws of xenosociology binding of all members of technological species for all time (\citet{GHAT1} calls this the ``monocultural fallacy'' and \citet{cirkovic18} calls such solutions ``exclusive.'') Indeed, many or most such proffered solutions ascribe universal behaviors to alien species that are not even universal among humans! 

And given how little we understand about the detectability of alien technology, it is not clear that the Fermi Paradox is a problem that even \textit{needs} a solution. Recall that Fermi posed his question in 1950, before the first SETI searches had even begun: it is therefore not about why we haven't found alien technology yet, it is about why they are not \textit{obviously on Earth today} \citep[see][and references therein.]{FermiParadox,Carroll19}

Similarly, work on ``hard steps,'' \citep{HardSteps} the Great Filter \citep{Filter}, the Great Silence \citep{Silence}, or the Eerie Silence \citep{Davies_Eerie} often presumes that there is a great puzzle to be solved, but to a large degree the puzzle derives from the rather unjustified assertion that if ``they'' existed, we would have noticed them by now. The validity of this axiom is far from obvious, and in my opinion it is false. Our completeness for our searches for alien technology is very low---indeed it has hardly even been calculated \citep[e.g.][]{Haystack}.

In my opinion, work in this direction is best focused on the problem of detectability, and on calculating what limits we can put on alien technology given the limited searching we have already done or could do.

\subsubsection{The Drake Equation}

The Drake Equation \citep{DrakeEquation} is another topic that generates a substantial stream of literature as authors rearrange, recompute, modify, reinterpret, and statistically analyze its various terms. 

The equation itself has served the field well as a heuristic for thinking about SETI and framing the entire field of astrobiology; indeed, each term could correspond to one of the units in an undergraduate Life in the Universe course. 

Quantitatively, it is best thought of as an elegant solution to a Fermi problem,\footnote{Fermi problems are estimation problems to be solved to an order of magnitude, the classic one being ``How many piano tuners are there in Chicago?'' The point is not to get the answer precisely, but to bound a seemingly impossible problem based on things you know roughly, such as the population of Chicago and what fraction of people own tuned pianos. Such exercises can greatly sharpen one's ability to hone in on the relevant aspects of a problem, and so is a favorite pastime among certain kinds of physicists. Answering the question by contacting the Chicago chapter of the Piano Technician's Guild would be considered cheating, or at best missing the point (indeed, most piano tuning likely takes place at music conservatories and is rendered by people doing the work part-time, two factors rarely considered in the classic solutions to that problem!).} and not as a fundamental equation of the field akin to the Schr\"odinger equation in quantum mechanics, or Newton's Laws of motion in classical mechanics. As such, critiques of SETI generally that focus on the Drake Equation miss the mark.

But the difficulty in using the equation to do more detailed analysis is no flaw: the equation was never meant to be \textit{solved}, it was meant to \textit{guide} us to a better understanding of the problem \citep{Drake1992}. That so many authors have critiqued, decried, repaired, reformulated, and extended the Drake Equation is a testament to its utility. Far from showing the limitations of the Drake's approach, such work actually vindicates it. 

But in the end, the Drake Equation's terms are fundamentally too uncertain and its assumptions to narrow for it to truly estimate the number of technological species in the Galaxy (or beyond), except to reiterate its original message: it is plausible that there could be a very large number of technological species in the Galaxy, which is a necessary condition for SETI. Such a conclusion was profound in 1961; today it hardly needs further reiteration or refinement.

So as with work on the Fermi Paradox, I suspect that further work on the Drake Equation will probably not provide significant new insight into how to search for technosignatures.

\subsection{Think About the Nine Axes}

\citeauthor{Sheikh2020}'s nine axes provide a framework for thinking about good SETI strategies, and were not intended to be objective or to produce an unambiguous ranking of strategies. The weights one assigns the axes when designing a figure of merit with them will depend strongly on one's priors on the kind and abundance of technosignatures and life in the universe. Nontheless, I think any SETI project should pay special attention to three of them in particular.

\subsubsection{Ancillary Benefits}

\label{sec:ancillary}
The primary purpose of SETI is, of coure, to make a positive detection or, barring that, to place an upper limit on a particular kind of technosignature. Secondarily, some searches produced significant ancillary benefits either inevitably or by design. For instance, the Breakthrough Listen backend at Green Bank observatory provides an outstanding platform for fast time domain radio astronomy of fast radio bursts and pulsars, and substantially contributes to those fields as a result \citep[e.g.][]{Michilli18,Zhang18,Gajjar18,Price19}. Similarly, the PANOSETI project promises new insights into fast optical transients \citep[e.g.][]{PANOSETI,Maire2019}, and the \^G project identified many extreme, anomalous objects of presumably natural origin \citep{GHAT3}.

This is consistent with Freeman Dyson's First Law of SETI Investigations: ``Every search for alien civilizations should be planned to give interesting results even when no aliens are discovered'' (F. Dyson, private communication). There are at least four good reasons for designing a SETI program to follow this advice. 

The first two are for the good of the field. The first is the political difficulty of needing to use null results to justify ever increasing budgets for ever more sensitive searches. Indeed, Dyson's primary concern was that all-or-nothing experiments did not have sustainable funding models: 
\begin{quote}It should be a fundamental principle in the planning of space operations, that every mission searching for evidence of life should have other exciting scientific objectives, so that the mission is worth flying whether or not it finds evidence of life. We should never repeat the mistake that was made with the Viking missions, whose advertised purpose was to give a definitive answer to the question whether life exists on Mars. After the Viking missions failed to find evidence of life, the further exploration of Mars was set back for 20 years. \citep{Dyson2003}
\end{quote}

The second is that if SETI is to exploit its many synergies with astrophysics and take its proper home under the astrobiology umbrella, it needs to avoid being cloistered, meaning its researchers should be valued participants in those broader disciplines, known for more than the upper limits they produce on technosignatures.

The other two reasons are for the SETI researchers personally. SETI is an underfunded field, and training new researchers in it should give them skills transferable to other parts of astronomy and beyond where better job prospects lie. 

And finally, as \citet{DrakeEquation} advised those that would follow him:
\begin{quote}
Our experience with Project Ozma showed that the constant acquisition of nothing but negative results can be discouraging. A scientist must have some flow of positive results, or his interest flags. Thus, any project aimed at the detection of intelligent extraterrestrial life should simultaneously conduct more conventional research. Perhaps time should be divided about equally between conventional research and the intelligent signal search. From our experience, this is the arrangement most likely to produce the quickest success.
\end{quote}

This point about ancillary benefits can be made too strongly, however. SETI is an important field that should stand on its own merits, and not only to the degree that it supports other kinds of astronomy. Indeed, many forms of SETI will primarily produce little to no data amenable significant astrophysical interpretation---most radio SETI data probably falls into this category. While it would be of course beneficial if those data had broader use, such searches have enough merit in other dimensions that they remain quite worthwhile. 

\subsubsection{Contrivance and Duration}

I also recommend counting the number of assumptions one is making on the ``contrivance/inevitability'' axis.  It can be easy to build a tower of assumptions about alien behavior that lead to a clean, unambiguous technosignature, and convince oneself that a search for that technosignature will likely succeed or else yield a strong null result. But tall towers teeter, and while it may be worthwhile exploring what is at the top, the utility of such a climb is often, in my experience, significantly less than the builder of the tower imagines.  A good strategy is to run one's assumptions by a SETI skeptic before building a tower too tall.

Finally, think about the duration axis in terms of $L$ in the Drake equation. Consider the number densities and rates of a given technosignature required for us to have any reasonable chance of observing short-lived phenomena with low duty cycles (for instance, cataclysmic events), and consider whether a search for such phenomena can ever produce useful or interesting upper limits. 

\subsection{Engage Experts}

SETI is closely related to other parts of astronomy. Especially when dealing with potential false positives---from natural confounders to terrestrial sources of interference---one works best as part of a team that includes people who know the instruments, data, and targets of the search best.  

This engagement also helps to grow the field. I have found that many astronomers and astrobiologists are happy to help with a SETI project, but do not know how they can contribute. Respecting their expertise is a great way to bring them in and improve one's science.

This is especially true when wandering outside of one's domain, for instance for an astrophysicist considering the nature of intelligence, the mathematical formulae for patterns of a species' expansion, the nuances of interspecies communication, the ethics of transmissions, or the effect on humanity of a clear SETI detection. These are well studied problems in the humanities and the life and social sciences, and while experts in those fields may not always have answers a physical scientist can immediately use for their purposes, they can help SETI practitioners bound such problems and prevent avoidable blunders. 

\subsection{Plan for Success, and for Null Results}

For observational programs, I strongly recommend having a fully formed plan that goes beyond simply searching for a given technosignature.  This means that one should plan for both success, and for a null result (I will not write ``failure,'' since a null result can be an important result \textit{per se} and not a failure at all.)   

Planning for success does not just mean planning for what to do after one is convinced that one has definitive signs of extraterrestrial technology (although one needs to have a good plan for that as well, beyond simply ``pop the champagne and call the press office''!).  It also means planning what to do between the detection of a stunning, interesting, intriguing, confusing, or even merely curious signal, and the eventual conclusion that alien life has been found. 

Searches should have detection thresholds tuned to generate a manageable number of candidate signals, and a robust follow-up program should be ready to pursue them. One should be prepared to document each signal and why it is or is not worthy of moving to the next stage of validation, and a pre-defined threshold for what constitutes sufficient evidence for a claim of a detection. 

It is possible, and perhaps likely, that the first evidence of extraterrestrial technology will not be from an unambiguous ``Wow!\ Signal'', but from a hard-to-understand ``Huh?\ Signal.'' In the latter case, it may take years or decades after the first public announcement of the signal before a consensus develops that alien technology has been detected.  Managing such announcements will take care, and practitioners should be prepared to solicit and heed the advice of experts in risk communication, sociology or anthropology, and science communication to help manage the announcement. After all, the announcement of evidence for the existence of \textit{technological} life may have significantly more amplified or different impact on the public than the announcements of potential biosignatures in the AH84001 meteorite or the clouds of Venus (both of which provide case studies that should be considered carefully.)

Planning for a null result is even more important, since that will be the more common outcome. Ideally, one would have calculated the significance of a null result \textit{before one begins searching} a given data set as part of the justification for looking in the first place.  Upper limits can be difficult to calculate rigorously (more difficult, often, than interpreting detections) but without such calculations a search has very little value to future researchers.  Injection-recovery methods are a particularly useful tool in this regard. 

Model-free anomaly searches, especially those that employ machine learning, may have a particularly difficult time describing what, exactly, it is that they did not find, and thus parameterize exactly which technosignatures they have shown do not exist.  But that problem is worth tackling.

\subsection{Use This Plan to Help Frame the Search}

Planning for success and null results is not just important for the progress of the field, but for the perception of the field, and therefore its future as a discipline.

 If SETI is seen merely as the practice of uncritically asking ``is it aliens?''\ about every anomaly, then it will be difficult to engage experts in useful data sets and leverage their expertise to advance the field, lest they become party to ``crying wolf'' at every poorly understood phenomenon. If the SETI literature is primarily a long string of null results and false alarms, then it will be seen as a risky endeavor, equivalent to buying a lottery ticket: something one might as well try, but not if it costs too much or will result in embarrassing headlines in the British tabloids. 

If, on the other hand, SETI develops a reputation for the rigorous calculation of upper limits and measurably advancing what we know is {\it not} out there, then its practitioners may be accepted as collaborators by even SETI skeptics (who, one would hope, would welcome such upper limits as confirming their priors). The goal should be to emulate the dark matter particle search community, which with each new instrument moves their upper limit thresholds lower and lower \citep[see][Appendix A]{Visions}. While they may never be able to rule out a particle of {\it any} cross section or mass, they can steadily shrink the remaining parameter space in which it can hide, provide clear markers of progress, and specifications for what the next useful experiment would look like. This can also help solve the funding paradox where it is the \textit{lack} of results that must justify ever increasing efforts to press on.

Or consider a planetary scientist with expertise in high resolution imagery or subsurface radar, who may look skeptically on a request to collaborate on a search for alien artifacts on the Moon or Mars. They might reasonably worry that they will be asked to co-author another ``Face on Mars'' paper when a strange rock formation inevitably triggers a machine learning algorithm or community scientist's interest \citep{Carlotto1997}. 

If, however, they understand that the goal is to rigorously and quantitatively put upper limits on the \textit{lack} of any such artifacts---while at the same time discovering new and interesting features that may or may not be such artifacts---then they are much more likely to appreciate the synergies of such collaboration, and to welcome abstracts and papers on the subject in their journals and at their conferences.

\subsection{Stay Broad-Minded}

We all come to the problem of SETI with very different priors for how SETI can succeed, and so will invariably encounter practitioners pursing what we feel are very unlikely or misguided paths to success. It helps to remember that the feeling may be mutual.

Avi Loeb and Cl\'ement Vidal have argued strongly that one should have very flat priors when considering natural versus artificial explanations for astrophysical phenomena.  Loeb in particular has been a strong public advocate for the hypotheses that a variety of phenomenena, from fast radio bursts to interstellar objects such as 1I/`Oumuamua, are evidence of alien technology.  

This at first glance would seem to be in violation of Carl Sagan's maxim that ``extraordinary claims require extraordinary evidence.'' But that maxim is essentially a paraphrasing of Bayes's Theorem: the prior that any particular technosignature will be found is presumably very small, and so requires a large amount of evidence to overcome.  Before applying the maxim to outr\'e claims, we must be careful to distinguish between the claim that our prior for a particular claim is too small, and the claim that the posterior probability after considering a piece of evidence is near one. 

Put another way, all great discoveries were, at some point, thought to be extremely unlikely, and the generation of unlikely hypothesis is thus an important, indeed essential, part of scientific progress. After all, part of the practice of science is being willing to update our priors regularly. Extraordinary and surprising objects or arguments can thus serve as case studies, and as opportunities for us to revisit those priors, even if those particular objects turn out in the end not to be examples of the phenomenon in question. For instance, my prior on the possibility of a prior indigenous technological species on Earth was once very small because I presumed such a species would have left traces that would be obvious to us today. After considering the matter and encountering the work of \citet{Schmidt19}, I realize that this is not the case, and that I have very little evidentiary basis for finding the hypothesis unlikely. My prior on this possibility is still low, but it is no longer very close to zero. \citep{PITS}

So while the ``broad priors'' stance can fairly be described as an ``aliens of the gaps'' philosophy, such an approach is a perfectly reasonable way to go about generating hypotheses in SETI, and we should welcome open-minded and creative papers that challenge our priors in the scientific literature \textit{provided that} they are are accompanied by sound scientific reasoning and, optimally, specific suggestions for how to test the hypothesis. It is also just generally good practice to respect the range of ``risk tolerance'' among practitioners in the field, and ensure that some fraction of our collective efforts are ``high-risk, high-reward.'' \citep{Loeb2010,Vickers2020}

Note that this is a separate question from how one should manage media attention to particular claims, particularly sensationalized stories in the yellow press. A discussion of that particular problem is beyond the scope of this work, but I will summarize my recommendation here by repeating: engage experts, and plan for success.

\subsection{Stay Skeptical, but not Cynical}

Not all SETI researchers believe they will have a good chance of success in their lifetimes, but such a belief surely animates much of the field. It can therefore be challenging to maintain a scientist's proper, healthy skepticism about one's own work, especially when coming across a particularly intriguing signal.

I suspect everyone who engages in the practice long enough will come across what looks to be a Wow!\ Signal and, at least briefly, dream of the success that will follow. The proper response to such a discovery is a stance of extreme skepticism: if one is not one's own harshest critic, one may end up embarrassing oneself, and losing credibility for the field. It is at these moments that Sagan's maxim should have its strongest force.

But one should also not let the product of such false alarms be a cynicism that leads one to focus entirely on upper limits and dismiss all candidate signals before they are thoroughly examined as just so much noise. There is a wonder that brought many of us into the field that must be nurtured and protected against the discouragement of years or decades of null results that \citeauthor{DrakeEquation} warned about. One should cherish each false alarm and ``Huh?\ signal" as an opportunity for hope and curiosity to flourish, ``till human voices wake us, and we drown.''\footnote{\citet{Prufrock}}

\acknowledgements{
I thank Jill Tarter, Adam Frank, Ravi Kopparaup, Jacob Haqq-Misra, Manasvi Lingam, Sofia Sheikh, and the students of the 2020 instance of Astro 576 at Penn State for many helpful discussions that helped shape this paper. I thank Adrian Lucy, David Grinspoon, and Stephen Wilson for help tracking down references. I thank the referees for helpful feedback.

I thank the participants of the Technoclimes Workshop and the KISS Workshop on Data-Driven Approaches to Searches for the Technosignatures of Advanced Civilizations for invigorating discussions on model-based versus anomaly-based detection algorithms, and inspiring Figure~\ref{fig:upperlimits}.  I thank my co-Is on an unsuccessful proposal, Joe Lazio, Adam Frank, Jacob Haqq-Misra, Matthew Povich, Paul Davies, Sara Walker, and Ravi Kopparapu, for their contributions to Figure~\ref{fig:chart}.

The Center for Exoplanets and Habitable Worlds and the Penn State Extraterrestrial Intelligence Center are supported by the Pennsylvania State University and its Eberly College of Science.

This research has made use of NASA's Astrophysics Data System Bibliographic Services. 

}

\newpage

\end{document}